\documentstyle[12pt,frascatiphys]{article}
\begin{document}
\title{ RELATIONSHIP BETWEEN THE QUARK CONDENSATE 
AND LOW-ENERGY $\pi\pi$ OBSERVABLES BEYOND $O(p^4)$\thanks{
 \,\,Work supported in part by the EEC-TMR Program, Contract N. CT98-0169 (EURODA$\Phi$NE).}. 
}

\author{L. Girlanda
 \\
{\em IFAE, Universitat Aut\`onoma de Barcelona, E-08193 Barcelona, Spain} } 
\maketitle
\baselineskip=14.5pt
\begin{abstract}
The two-flavor Gell-Mann--Oakes--Renner ratio is expressed in terms of low-energy $\pi\pi$ observables including the $O(p^6)$ double chiral logarithms, computed in Generalized Chiral Perturbation Theory. It is found that their contribution is important and tends to compensate the one from the single chiral logarithms.
\end{abstract}
\baselineskip=17pt
\section{Introduction}
Low-energy $\pi\pi$ scattering offers the rare possibility to test a fundamental property of the QCD vacuum,  the strength of quark-antiquark condensation $\langle \bar q q \rangle$\cite{gchpt}. The knowledge of this quantity is crucial to understand the mechanism of spontaneous chiral symmetry breakdown (SB$\chi$S) in the Standard Model.
The quark condensate, as every other order parameter, depends on the number of massless flavors $N_f$, and will in general experience a paramagnetic suppression as $N_f$ increases\cite{para}.
It turns out that what can actually be measured from low-energy $\pi\pi$ scattering is the quark condensate in the limit of two massless flavors, $N_f=2$.  The question to be addressed is to which extent $N_f=2$ is close to a critical point where $\langle \bar q q \rangle$ disappears and chiral symmetry is eventually broken by higher-dimensional order parameters\cite{stern}.
The proper framework to analyze phenomenologically the issue is provided by the generalized version of Chiral Perturbation Theory ($\chi$PT)\cite{gchpt} in the case of SU(2)$\times$SU(2) chiral symmetry.
This is a reorganization of the standard expansion of $\chi$PT\cite{gale} in which the quark condensate parameter $B$ is considered formally as a small quantity, in order to account for the possibility that the linear and quadratic terms, in the expansion of the pion mass  $M_\pi^2$ in powers of quark masses $m_u = m_d = \hat m$, be of the same order.
The chiral counting is modified accordingly,
\begin{equation}
B \sim \hat m \sim O(p), \hspace{1cm}{\cal L}= \tilde {\cal L}^{(2)} + \tilde {\cal L}^{(3)} + \tilde {\cal L}^{(4)} + \ldots
\end{equation}
with  $\tilde {\cal L}^{(d)}$ containing additional terms, relegated  in higher orders by the standard counting.
The complete effective Lagrangian up to $O(p^4)$ can be found in Refs.\cite{su2gchpt}, together with its renormalization at 1~loop.

At present, the low-energy $\pi\pi$ phase-shifts are rather poorly known, but considerable improvements are expected to come soon from new high-luminosity $K_{e4}$ decays experiments, performed at Brookhaven and DA$\Phi$NE.
These new data, together with the recent numerical solutions of Roy Equations\cite{roysol}, will allow to extract the two $S$-wave scattering lengths or, equivalently, the two parameters $\alpha$ and $\beta$, introduced in Ref.\cite{pipi1} and representing respectively the amplitude and the slope at the symmetrical point $s=t=u=4/3 M_\pi^2$.
In this perspective, we have established the relationship between the two-flavor quark condensate, expressed through the deviation from the Gell-Mann--Oakes--Renner relation,
  
\begin{equation} \label{eq:gor}
x_2^{\mathrm{GOR}}= \frac{| 2 \hat m \langle \bar q q \rangle |}{F_\pi^2 M_\pi^2},
\end{equation}
and the parameters $\alpha$ and $\beta$, including the leading $O(p^6)$ double logarithmic corrections to the 1-loop result of Generalized $\chi$PT.

\section{Double chiral logs}
Due to the smallness of the pion mass, double chiral logarithms are among the potentially most dangerous contributions at order $O(p^6)$. As first pointed out in Ref.\cite{bellucci} they can be obtained from a 1-loop calculation, using the fact that, in the renormalization procedure, non-local divergences must cancel. Setting the space-time dimension $d=4+\omega$ to regulate the theory, all the low-energy constants (l.e.c.'s) of the generalized Lagrangian with $k$ derivatives and $n$ powers of the scalar source, $c_{(k,n)}$, have dimension $2-k-n$, except $F^2$ which has dimension $[F^2]=d-2$. We thus replace $F^2$ with $\mu^{2 \omega} F^2$, making appear explicitly the scale parameter $\mu$ brought in by the regularization procedure.
Since each loop involves a factor $F^{-2}$, the chiral expansion of a generic amplitude ${\cal A}$, apart from an overall dimensional factor, takes the form,
\begin{equation}
 {\cal A} \sim {\cal A}_{\mathrm{tree}} + \left( \frac{M_\pi}{\mu}\right)^{\omega} \sum_i P^{(1)}_i (c_{(k,n)}) g_i^{\mathrm{1-loop}}  + \left(\frac{M_\pi}{\mu} \right)^{2 \omega} \sum_i P^{(2)}_i (c_{(k,n)}) f_i^{\mathrm{2-loop}} + \ldots,
\end{equation}
where $P_i^{1,2}$ are polynomials in the l.e.c.'s
and $g_i$ and $f_i$ are loop-functions of the kinematical variables, expressed in terms of dimensionless quantities.
After writing the Laurent expansions of the loop-functions and of the coupling constants,
\begin{equation}
f_i=\frac{f_{i,2}}{\omega^2} + \frac{f_{i,1}}{\omega} + f_{i,0} + \ldots, \hspace{1cm}
g_i=\frac{g_{i,1}}{\omega} + g_{i,0} +  \ldots, \hspace{1cm}
c_i=\frac{\delta_i}{\omega} + c_i^r + \ldots,
\end{equation}
and imposing the cancellation of the non-local divergences $\sim 1/\omega \log M_\pi$, one finds that the double chiral logarithms are given by the residues $g_{i,1}$ of the pole in $\omega$, and always occur in the same combination with the terms $\sim c_i^r \log (M_\pi/\mu)$,
\begin{equation}
{\cal A}^{\mathrm{llogs}} \sim \frac{g_{i,1}}{8} \left[ - \frac{\Gamma_i}{16 \pi^2} \log \frac{M_\pi^2}{\mu^2} + 4 c_i^r \right] \log \frac{M_\pi^2}{\mu^2},
\end{equation}
$\Gamma_i$ being the $\beta$-function coefficients of the l.e.c. $c_i$. Notice that at order $O(p^6)$ we never have to deal with products like $\Gamma_i \Gamma_j$, since all 1-loop divergences are at least $O(p^4)$.
We display the result for $M_{\pi}$ and $F_{\pi}$, where all l.e.c.'s, here and in the following, are renormalized at a scale $\mu$:
\begin{eqnarray}
\frac{F_\pi^2}{F^2} M_{\pi}^2 &=& 2 B \hat m + 4 A \hat m^2 
 + \left( 9 \rho_1 + \rho_2 + 20 \rho_4 + 2 \rho_5 \right) \hat
m^3  \nonumber \\ 
&& + \left( 16 e_1 + 4 e_2 + 32 f_1 + 40 f_2  + 8 f_3 + 96 f_4
\right) \hat m^4  \nonumber \\
&& + 4 a_3 M_{\pi}^2 \hat m^2 -
 \frac{M_{\pi}^2}{32 \pi^2 F_{\pi}^2}  \left( 3 M_{\pi}^2 + 20 A \hat m^2
 \right) \log \frac{M_{\pi}^2}{\mu^2} \nonumber \\
&& + \left[ \frac{33}{8} +\frac{65}{2} \frac{A \hat m^2}{M_{\pi}^2} + 60
\left( \frac{A \hat m^2}{M_{\pi}^2} \right)^2 \right]  M_{\pi}^2 \left( \frac{M_\pi^2}{16
\pi^2 F_\pi^2} \log \frac{M_{\pi}^2}{\mu^2}
\right)^2  , \\
F_{\pi}^2 &=& F^2 \left[ 1 + 2 \xi^{(2)} \hat m + \left( 2 a_1 + a_2 +
4 a_3 + 2 b_1 - 2 b_2 \right) \hat m^2- \frac{M_{\pi}^2}{8 \pi^2
F_{\pi}^2} \log \frac{M_{\pi}^2}{\mu^2}\right. \nonumber \\
&&\left. + \frac{7}{2}
\frac{M_{\pi}^4}{F_{\pi}^4} \left( \frac{1}{16 \pi^2} \log \frac{M_{\pi}^2}{\mu^2}
\right)^2   \right]. 
 \end{eqnarray}
\section{The GOR ratio}
The amplitude for $\pi\pi$ scattering up to $O(p^6)$ has been first given in Ref.\cite{pipi1}, using dispersive techniques, independently of any assumption about the size of the chiral condensate. It can be expressed in terms of 6 parameters, $\alpha,\beta,\lambda_1,...,\lambda_4$,
\begin{equation} \label{ampl}
 A(s|t,u) = A_{\mathrm{KMSF}}(s|t,u;\alpha,\beta;\lambda_1, \lambda_2,
 \lambda_3,
 \lambda_4) +  O \left[ \left({p \over \Lambda_{\mathrm{H}}} \right)^8
 , \left( {M_{\pi} \over
 \Lambda_{\mathrm{H}}} \right)^8 \right].
 \end{equation}        
The $\lambda_i$'s can be determined from a set of twice subtracted fixed-$t$ dispersion relations (Roy Equations), whereas $\alpha$ and $\beta$ can be related to the two subtraction constants.
Most of the sensitivity to $\langle \bar q q \rangle$ is contained in the parameter $\alpha$, which, at tree level, varies from 1 to~4 if $\langle \bar q q \rangle$ is decreased from its standard value down to zero.
The complete two-loop  S$\chi$PT calculation of Ref.\cite{loropipi} allows in addition to express the six parameters in terms of the l.e.c.'s\cite{l1l2}. It is interesting to notice that, in this standard case,  the double chiral logarithms constitute by far the largest $O(p^6)$ contribution to $\alpha$.
An explicit calculation in G$\chi$PT, along the lines described in the previous section, yields, 
\begin{eqnarray}
\alpha &=& \frac{F^2}{F_\pi^2 M_\pi^2} \biggl[2 B \hat m + 16 A \hat m^2 -
4 M_{\pi}^2 \xi^{(2)} \hat m  
+ \left( 81 \rho_1 + \rho_2 + 164 \rho_4 + 2 \rho_5 \right) \hat
m^3  \nonumber \\
&& - 8 M_{\pi}^2 \left( 2 b_1 - 2 b_2 - a_3 - 4 c_1 \right) \hat m^2 
\nonumber \\
&& + 16 \left( 6 A a_3 + 16 e_1 + e_2 + 32 f_1 + 34 f_2 + 2 f_3 +
72 f_4 \right) \hat m^4  \nonumber \\
&& - \frac{M_{\pi}^2}{32 \pi^2 F_{\pi}^2} \left( 4 M_{\pi}^2 + 204 A \hat m^2  + 528 \frac{ A^2 \hat
m^4}{M_{\pi}^2} \right) \log \frac{M_{\pi}^2}{\mu^2}  \nonumber \\
&& - \frac{ 1}{32 \pi^2 F_{\pi}^2} \left[ M_{\pi}^4 + 88 A \hat m^2
M_{\pi}^2 + 528 A^2 \hat m^4 \right] \nonumber \\
&& + M_{\pi}^2 \left[ \frac{533}{72}  + \frac{18817}{30} \frac{A \hat
m^2}{M_{\pi}^2} +
\frac{61076}{15} \left( \frac{A \hat m ^2}{M_{\pi}^2} \right)^2 \right.
\nonumber \\
&& \left. + 5808
\left( \frac{A \hat m ^2}{M_{\pi}^2} \right)^3 \right] \left( \frac{M_{\pi}^2}{16 \pi^2 F_{\pi}^2}
\log \frac{M_{\pi}^2}{\mu^2} \right)^2 \biggr], \\
\beta &=& 1 + 2 \xi^{(2)} \hat m - 4 {\xi^{(2)}}^2 \hat m^2 + 2 \left( 3 a_2
+ 2 a_3 + 4 b_1 + 2 b_2 + 4 c_1 \right) \hat m^2  \nonumber  \\
&& - \frac{4 M_{\pi}^2}{32 \pi^2 F_{\pi}^2} \left( 1 + 10 \frac{ A \hat
m^2}{M_{\pi}^2 } \right) \left( \log \frac{M_{\pi}^2}{\mu^2} + 1 \right)
\nonumber \\
&& + \left( \frac{151}{36} M_{\pi}^4 + \frac{400}{3} M_{\pi}^2 A \hat m^2 +
420 A^2 \hat m^4 \right) \left[  \frac{1}{16 \pi^2 F_{\pi}^2}
\log \frac{M_{\pi}^2}{\mu^2} \right]^2 , \\
\lambda_1 &=& 2  l_1 - \frac{1}{48 \pi^2} \log
\frac{M_{\pi}^2}{\mu^2}  - \frac{ 1}{ 36 \pi^2}
+ \left( \frac{25}{18} + \frac{130}{9} \frac{A \hat m^2}{M_{\pi}^2}
\right) \left[  \frac{M_{\pi}^2}{16 \pi^2 F_{\pi}^2}
\log \frac{M_{\pi}^2}{\mu^2} \right]^2 , \\
\lambda_2 &=&  l_2 - \frac{1}{48 \pi^2} \log
\frac{M_{\pi}^2}{\mu^2}  - \frac{ 5}{ 288 \pi^2} 
+ \left( \frac{5}{3} + \frac{80}{9} \frac{A \hat m^2}{M_{\pi}^2} \right)
\left[  \frac{M_{\pi}^2}{16 \pi^2 F_{\pi}^2}
\log \frac{M_{\pi}^2}{\mu^2} \right]^2 , \\
\lambda_3 &=& \frac{10}{9} \left[ \frac{1}{16 \pi^2} \log
\frac{M_{\pi}^2}{\mu^2} \right]^2, \hspace{1cm} \lambda_4 = -\frac{5}{18} \left[ \frac{1}{16 \pi^2} \log
\frac{M_{\pi}^2}{\mu^2} \right]^2.
\end{eqnarray}
It is easy to check that these formulae, when restricted to the standard case,  agree with the ones displayed in Ref.\cite{l1l2} based on the complete two-loop calculation.
\begin{figure}[h]
 \vspace{9.0cm}
\includegraphics{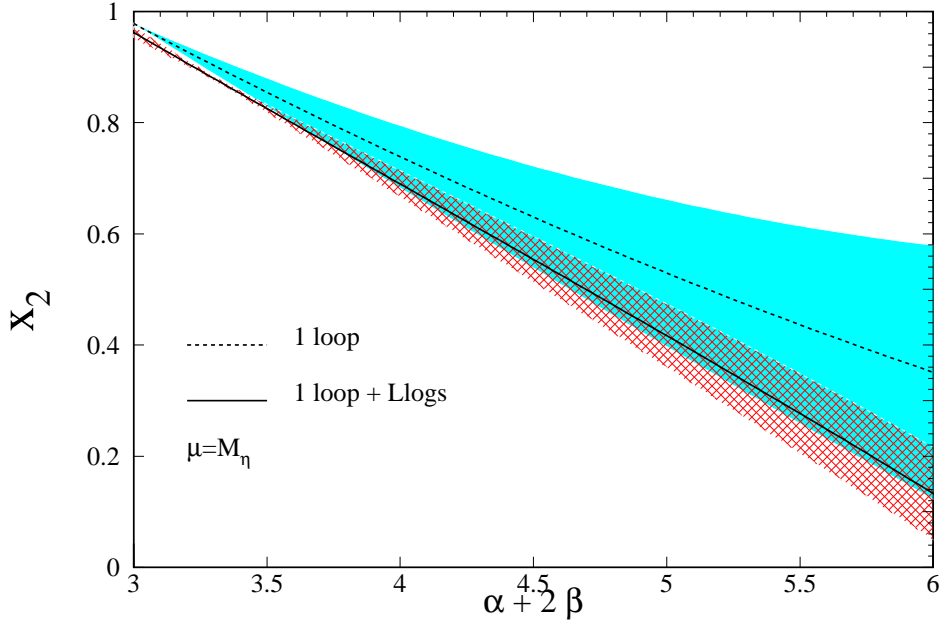}
 \caption{\it The GOR ratio at 1 loop including (continuous) and not including (dashed) the double logarithms. The scale is set to $\mu=M_\eta \pm 250$~MeV. 
    \label{fig:gor} }
\end{figure}
Eliminating the constant $A$ in favor of $\alpha$ and  $\xi^{(2)}$ in favor of $\beta$, one can express the GOR ratio of Eq.~(\ref{eq:gor}) as function of the combination $\alpha + 2 \beta$,
\begin{eqnarray}
x_2^{\mathrm{GOR}} & = &  \frac{2 \hat m B F^2}{F_{\pi}^2 M_{\pi}^2} = 2 -
\frac{ \alpha + 2 \beta}{3} + \frac{F^2}{F_{\pi}^2 M_{\pi}^2} \left( 15
\rho_1^{(2)} - \rho_2^{(2)} + 28 \rho_4^{(2)} - 2 \rho_5^{(2)} \right) \hat
m^3  \nonumber \\
&& + \left[ 4 a_2 +   8 \left( \frac{\alpha + 2 \beta }{3} - 1  \right) a_3
+ 8 b_2 + 16 c_1 \right] 
\hat m^2 \nonumber \\
&& + \frac{64}{M_{\pi}^2} ( e_1 + 2 f_1 + 2 f_2 + 4 f_4 ) \hat m^4 
+ \frac{M_{\pi}^2}{288 \pi^2 F_{\pi}^2} \left(
\alpha + 2 \beta \right) \left[ 24 - 11 ( \alpha + 2 \beta) \right]
\nonumber \\
&& +
 \left[ 6 + \frac{5}{3} (\alpha + 2 \beta) - \frac{11}{9} (\alpha + 2
\beta)^2 \right] \frac{M_{\pi}^2}{32 \pi^2 F_{\pi}^2} \log
\frac{M_{\pi}^2}{\mu^2} +\left[ \frac{11}{60} -
\frac{4169}{1080} (\alpha + 2 \beta) \right. \nonumber \\
&& \left. + \frac{5639}{1620} (\alpha + 2 \beta)^2
- \frac{121}{108} (\alpha + 2 \beta)^3 \right] \left( \frac{M_{\pi}^2}{16 \pi^2 F_{\pi}^2} \log
\frac{M_{\pi}^2}{\mu^2} \right)^2  . 
\end{eqnarray}
Fig.~\ref{fig:gor} shows this  function  for  $\mu=M_\eta$. The upper left corner represents the standard case, corresponding to $\alpha\sim 1, \beta\sim 1$, while higher values for $\alpha + 2 \beta$ would imply a significant departure from that picture.
The dashed line is the result without including the $O(p^6)$ double logarithms. The bands are obtained varying the $\chi$PT scale by $\pm$250~MeV and treating the unknown l.e.c.'s as randomly distributed around zero with magnitude according to na\"{\i}ve dimensional analysis,
\begin{equation}
\rho_i \sim \frac{1}{\Lambda_{\mathrm{H}}}, \hspace{2cm} a_i,b_i,c_i,e_i,f_i \sim  \frac{1}{\Lambda_{\mathrm{H}}^2}.
\end{equation}
 Most of the uncertainty comes from the $\mu$-dependence, which however, quite interestingly, almost cancels when including the double logarithms. The contribution from the latters is found to be rather large, although smaller that the $O(p^4)$ one, and tends to compensate for the 1-loop shift. An additional uncertainty (of difficult estimation), from the remaining $O(p^5,p^6)$ pieces, should be understood in Fig.~\ref{fig:gor}.

\begin{thebibliography}{99}

\bibitem{gchpt} N.H. Fuchs, H. Sazdjian and J. Stern, Phys. Lett. {\bf B269} (1991), 183;
Phys. Rev.  {\bf D47} (1993), 3814;
\\
M. Knecht, B. Moussallam and J. Stern, in The Second Da$\Phi$ne Physics
Handbook, eds. L.~Maiani, G.~Pancheri and N.~Paver, 1995; hep-ph/9411259.

\bibitem{para} B. Moussallam, Eur. Phys. J. {\bf C14} (2000), 111
; hep-ph/0005245.
\\
S.~Descotes, L.~Girlanda and J.~Stern, JHEP {\bf 01} (2000), 41.

\bibitem{stern} J. Stern, hep-ph/9801282.

\bibitem{gale}J.~Gasser and H.~Leutwyler, Ann. Phys. (NY)\ {\bf 158}
 (1984), 142; 
Nucl. Phys. {\bf B250} (1985), 465.

\bibitem{su2gchpt} L. Ametller, J. Kambor, M. Knecht and P. Talavera, Phys. Rev. {\bf D60} (1999), 094003.
\\L. Girlanda and J. Stern, Nucl. Phys. {\bf B575} (2000), 285.

\bibitem{roysol} B.~Ananthanarayan, G.~Colangelo, J.~Gasser and H.~Leutwyler,
hep-ph/0005297.

\bibitem{pipi1}M.~Knecht, B.~Moussallam, J.~Stern and N.H.~Fuchs, Nucl. Phys. {\bf B457} (1995), 513; 
Nucl. Phys. {\bf B471} (1996), 445.              
                                            
\bibitem{bellucci} S. Bellucci, J. Gasser and M.E. Sainio, Nucl. Phys. {\bf B423} (1994), 80.



\bibitem{loropipi}J.~Bijnens, G.~Colangelo, G.~Ecker, J.~Gasser and M.~Sainio,
Phys. Lett. {\bf B374} (1996), 210;
 Nucl. Phys. {\bf B508} (1997), 263.

\bibitem{l1l2}L.~Girlanda, M.~Knecht, B.~Moussallam and J.~Stern, Phys. Lett. {\bf B409} (1997), 461.

               
\end{thebibliography}
\end{document}